# HOW FAR AWAY ARE GAMMA–RAY BURSTERS?

Bohdan Paczyński

Princeton University Observatory, Princeton, NJ 08544–1001

Visiting Scientist, National Astronomical Observatory, Mitaka, Tokyo, 181, Japan

E-mail: bp@astro.princeton.edu

## ABSTRACT

The positions of over 1000 gamma-ray bursts detected with the BATSE experiment on board of the Compton Gamma Ray Observatory are uniformly and randomly distributed in the sky, with no significant concentration to the galactic plane or to the galactic center. The strong gamma-ray bursts have an intensity distribution consistent with a number density independent of distance in Euclidean space. Weak gamma-ray bursts are relatively rare, indicating that either their number density is reduced at large distances or that the space in which they are distributed is non-Euclidean. In other words, we appear to be at the center of a spherical and bounded distribution of bursters. This is consistent with the distribution of all objects that are known to be at cosmological distances (like galaxies and quasars), but inconsistent with the distribution of any objects which are known to be in our galaxy (like stars and globular clusters). If the bursters are at cosmological distances then the weakest bursts should be redshifted, i.e. on average their durations should be longer and their spectra should be softer than the corresponding quantities for the strong bursts. There is some evidence for both effects in the BATSE data.

At this time the cosmological distance scale is strongly favored over the galactic one, but is not proven. A definite proof (or dis-proof) could be provided with the results of a search for very weak bursts in the Andromeda galaxy (M31) with an instrument $\sim 10$ times more sensitive than BATSE.

If the bursters are indeed at cosmological distances then they are the most luminous sources of electromagnetic radiation known in the universe. At this time we have no clue as to their nature, even though well over a hundred suggestions were published in the scientific journals. An experiment providing $\sim 1$ arc second positions would greatly improve the likelihood that counterparts of gamma-ray bursters are finally found. A new interplanetary network would offer the best opportunity.

*Subject headings:* cosmology – gamma-ray bursts

## 1. INTRODUCTION

The title of this debate is: "The Distance Scale to Gamma-Ray Bursts". However, the real issue is: how do we do science, how do we reason, and what reasoning do we find convincing? My starting point is the statement: we do not know what gamma-ray bursters are and what makes them burst. Given this ignorance I shall present a standard astronomical approach to the distance determination. At the end I shall come with what I consider a good case for the distance being cosmological, but no clue whatsoever about the nature of the sources.

I think the majority of advocates of the galactic origin of the bursts assumed from the beginning that the sources are related to some neutron stars, and the distance follows from this assumption. Within this paradigm the distance to the bursters used to be $\sim 100$ parsecs just a few years ago, but it has increased by three orders of magnitude to $\sim 100$ kiloparsecs today (Lamb 1995). Following Hakkila et al. (1994) I shall use the term 'galactic corona' rather than 'galactic halo' for the proposed population of gamma-ray bursters at $\sim 100$ kpc. The reason is that the required distribution is unlike any observed or theoretically proposed galactic halo.

The information about the observed properties of gamma-ray bursts was provided by Fishman (1995) in his introductory paper. I shall summarize what I consider to be the most important for their distance scale. The observed distribution of gamma-ray bursts appears to be isotropic. It is true that a number of researchers found small departures from perfectly uniform distribution by selecting various sub-samples of the bursts (e.g. Quashnock & Lamb 1993), but these small irregularities were not confirmed by the new data, and even the correctness of the original claim is in doubt (Rutledge & Lewin 1993). It is natural that any finite sample must exhibit some statistical fluctuations. So far no specific suggestion of anisotropy survived the test of time.

It is well known that if sources of any kind are uniformly distributed in a 3-dimensional Euclidean space then the number of sources, $N$, that appear to be brighter than some limit $F$ is proportional to that limit to the power $-1.5$, that is we have: $N \sim F^{-1.5}$. This relation is observed to be approximately satisfied by the bright stars as well as by the bright galaxies, even though the nature of the two types of objects, as well as their distances, are vastly different.

The best statistics of bright gamma-ray bursts is provided by the instrument PVO – Pioneer Venus Orbiter, which was collecting information for 13 years, and registered over 200 strong bursts (Fenimore et al. 1992). The distribution of these bursts is well approximated with the relation $N \sim F^{-1.4}$, indicating that the sources of apparently strong, and presumably relatively nearby bursts have a uniform number density out to some distance. The most sensitive instrument to date is BATSE, roughly 30 times more sensitive than PVO. The overall distribution of BATSE bursts may be approximated with the relation $N \sim F^{-0.8}$ (Meegan 1992), indicating that the number of apparently weak, and presumably distant bursts is relatively small, as if their number density were reduced at large distances, or if space at some large distance were no longer Euclidean. The distribution of burst intensities detected with instruments which had a sensitivity between PVO and BATSE can be approximated with a power law with a slope which is between –1.4 and –0.8 (Tamblyn & Melia 1993). In fact, when the PVO and BATSE bursts are combined, a gradual transition from a slope close to –1.5 on the bright end to –0.8 at the faint end becomes apparent (Fenimore et al. 1993). It appears that the results from all instruments provide a consistent view: the relatively nearby bursts are approximately uniformly distributed with distance, but beyond some distance the number density of bursters decreases, or space becomes non-Euclidean.

It is very important to consider the two results together: the isotropic distribution in the sky with the apparent shortage of weak bursts in all directions. The conclusion is: we appear to be at the center, or near the center of a spherically symmetric and bounded distribution of gamma-ray bursters.

Another very important feature of a gamma-ray burst is its spectrum. The spectra are very broad, covering many decades of photon energies, extending in extreme cases to photons as soft as 1 keV, and in some cases to photons as hard as 18 GeV (Hurley et al. 1994). A typical spectrum may be approximated with a broken power law (Schaefer et al. 1992, 1994, Band et al. 1993), and it is so broad that there is no doubt it is non-thermal. I think there is a consensus on this issue.

## 2. DISTANCE DETERMINATION IN ASTRONOMY

It is useful to recall how the distances are measured for various astronomical objects. This information can be found in any standard textbook. The simplest and the most direct method is the trigonometric parallax: a stellar position in the sky varies while the earth orbits the sun. With modern instruments stellar distances can be measured this way out to $\sim 100$ parsecs. This is a purely geometrical method.

A so called dynamical parallax is applicable to binaries for which spectroscopic orbits can be combined with the astrometric orbits. The geometry of orbital motion is measured in two ways, with spectroscopy and astrometry providing us with the linear and angular dimensions, respectively, and their ratio gives the distance to the binary.

Another purely geometrical method uses the apparent proper motions and radial velocities of a group of stars which cover a large area in the sky, like the Hyades cluster. A combination of the observed range of proper motions and radial velocities across the moving group allows the distance to be measured.

A purely geometrical method which works on a truly cosmological scale is based on gravitational lensing: the two images of a distant quasar are seen along the two different paths, and the light travel time is different along the two. If the brightness of the lensed quasar varies then the time lag between the observed variation of the two images is equal to the difference in the two path lengths divided by the speed of light. Other things being equal the time delay is proportional to the distance.

There are many simple but indirect geometrical methods known. If a new object appears to be associated with one with a known distance, then the distance to the new source becomes known as well. An example is a BL Lac type object with a featureless spectrum observed to be at the center of a galaxy with known redshift.

If a source is found behind a known object then its distance must be larger. An example is another BL Lac object with absorption lines in its spectrum observed to be at the same redshift as a galaxy which is seen in the same direction: the BL Lac object must be behind the galaxy.

If we can identify a known type of object or event, like a cepheid or a supernova of Type Ia, in a remote galaxy we may use the known absolute magnitude of our "standard candle" to measure the distance.

A powerful method of distance determination is based on the observed distribution of sources — this is another purely geometrical approach, described in the next section.

The only physical method which works well deals with thermal sources that are optically thick, and therefore they radiate like a black body, with a somewhat distorted spectrum. The observed spectrum is used to estimate the surface temperature, and hence the surface brightness. The radial velocity variations in an eclipsing binary, or in a pulsating star, are used to measure the linear size of the star, and hence its area. The product of the area and the surface brightness is the intrinsic (absolute) stellar luminosity, while the apparent luminosity (the flux) can be measured directly. The ratio of the two is proportional to the square of the distance. A popular technique of this kind is known as the Baade-Wesselink method.

Many other methods exist, but they are usually related to those which have been described. Inspecting a long list of proven methods we find there is only one which is applicable to gamma-ray bursters: we have to use their observed distribution, as described in the following section. Lamb (1995) uses another method related to those listed above: a hypothetical connection between the gamma-ray bursts and the soft gamma repeaters, to which the distance is known. The analysis of that approach is given in section 4.

## 3. DISTRIBUTIONS OF VARIOUS OBJECTS IN THE SKY

There are billions of stars, galaxies and other objects in the sky. There are hundreds of different types of objects known to astronomers. But there are only a few distinctly different types of distributions known. Almost every year a new class of objects is found. But a fundamentally new distribution has not been found in decades.

All objects known to exist in the inner solar system are strongly concentrated towards the ecliptic: the orbits of all planets, asteroids, zodiacal dust, comets in the Kuiper belt. At larger distances, between our inner solar system and the closest stars, the only known objects are the comets in the Oort cloud. Their distribution is almost spherical, but not exactly: the shape is affected by the tidal disturbances of our galaxy (Clarke et al. 1994). If the bursts came from the Oort cloud, then we should have seen by now the effect of these tidal distortions in the sky distribution of the bursts, and we have not. The Oort cloud comets have yet another property that makes them unacceptable as candidates for gamma-ray bursts: their number density varies with distance, and it is not uniform anywhere.

Once we get out of the solar system we encounter nearby stars, which are distributed more or less uniformly in space, apart from their tendency to 'cluster' in binaries and multiple systems. The distribution over the sky is almost isotropic, as shown in Fig. 1. Some apparent clustering is due to non-uniformity of the search procedure.

When the more distant stars are placed in a sky map their tendency to concentrate near the galactic plane becomes apparent: since Galileo we know that the Milky Way is made of stars. Among distant objects which are readily detected across the galaxy and which are related to stars of moderate mass are the planetary nebulae, the remnants of old red supergiants. A few thousand years ago the extended envelopes of those former supergiants flew away from their hosts at $\sim 20$ km/sec, and formed ring-like nebulae which are named "planetary". The ones older than a few thousand years expanded so much that they are too diffuse to see. The nuclei of the nebulae, former cores of red supergiants, are on their way to becoming white dwarfs. Planetary nebulae are very bright, and readily discovered out to large distances. Their concentration to the galactic plane demonstrates that they belong to the galaxy (cf. Figure 2).

There are also galactic halo objects in the astronomical inventory. The most extreme case of a halo distribution known to date is offered by globular clusters. These are the oldest components of our galaxy, with nearly spherical distribution, reaching out to tens of kiloparsecs. Figure 3 shows their positions in the sky: a very strong concentration to the galactic center is striking. This is a general property of all known halo objects. Notice that globular clusters are not distributed spherically around us, but they are almost spherically distributed around the galactic center. As we are 8 kiloparsecs away from the galactic center, we see them concentrated in that direction.

All these galactic objects: globular clusters, planetary nebulae and stars, have been detected in the nearby galaxies. This is a general property of all known galactic objects: none of them is specific to our galaxy only, all of them are also found in other galaxies, provided they are luminous enough to be detectable.

The distribution of nearby galaxies is shown in Fig. 4. Their clustering is very strong. In fact, almost all nearby galaxies are members of the Virgo cluster. Looking farther out we find that the more distant galaxies are still clustered but their overall distribution becomes more and more uniform. However, no optical or infrared survey has been deep enough to demonstrate directly that the distribution becomes truly homogeneous and isotropic. This is possible with radio surveys which detect very luminous and very rare radio galaxies and quasars. Because of their enormous radio power these objects are detected at truly cosmological distances, and even the brightest sources show an isotropic and random distribution over the sky, as presented in Fig. 5.

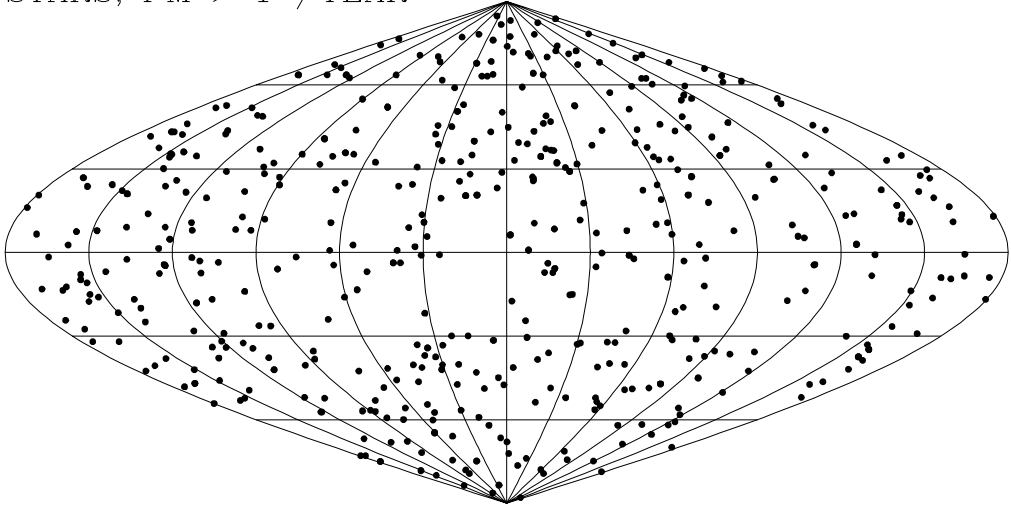

Fig. 1.— The distribution of 528 nearby stars is shown in galactic coordinates. The data were obtained from the CDS in Strasbourg (I/87A) and it is based on W. J. Luyten (1976). These are stars with a proper motion in excess of one second of arc per year. The distribution is approximately isotropic and random. Some apparent clustering is due to non-uniform sky coverage of the search for high proper motion stars.

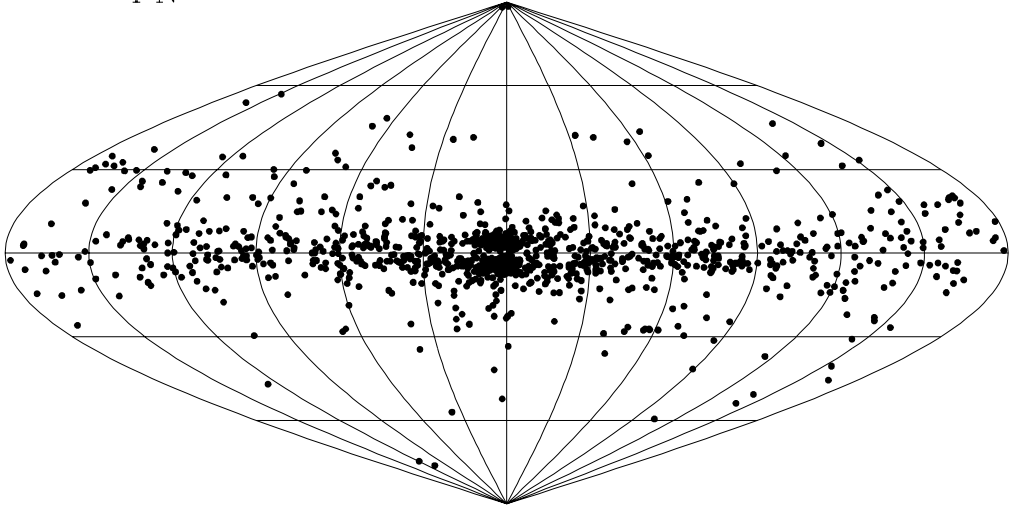

Fig. 2.— The distribution of the 1143 galactic planetary nebulae (PN) is shown in galactic coordinates. The data were obtained from the CDS in Strasbourg (V/84) and it is based on A. Acker et al. (1992, Strasbourg-ESO Catalogue of Galactic Planetary Nebulae). Notice the strong concentration of objects toward the galactic plane, typical for the distribution of distant stars.

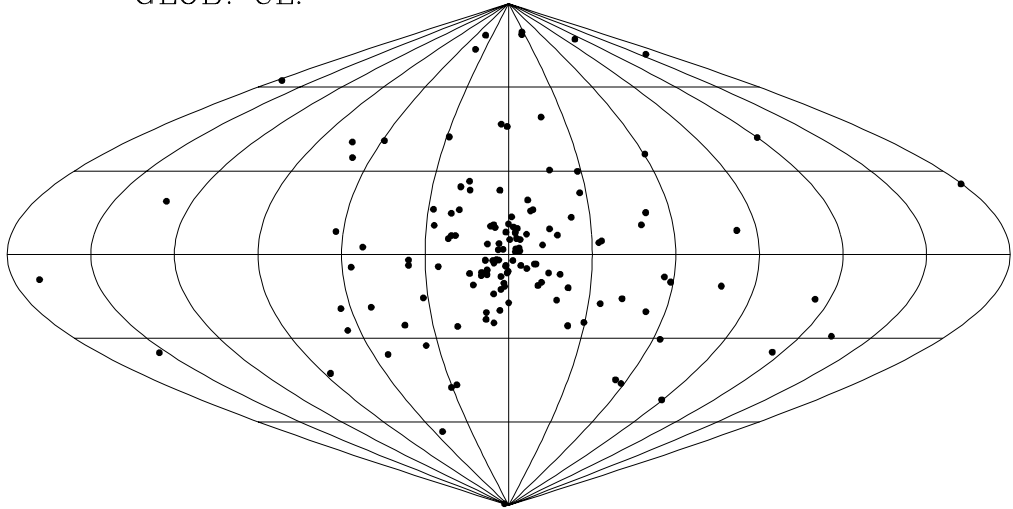

Fig. 3.— The distribution of the 160 galactic globular clusters (GLOB. CL.) is shown in galactic coordinates. The data were obtained from the CDS in Strasbourg (VII/103) and it is based on R. Monella (1985, Coelum LIII, 287). Notice the strong concentration toward the galactic center of these typical galactic halo objects.

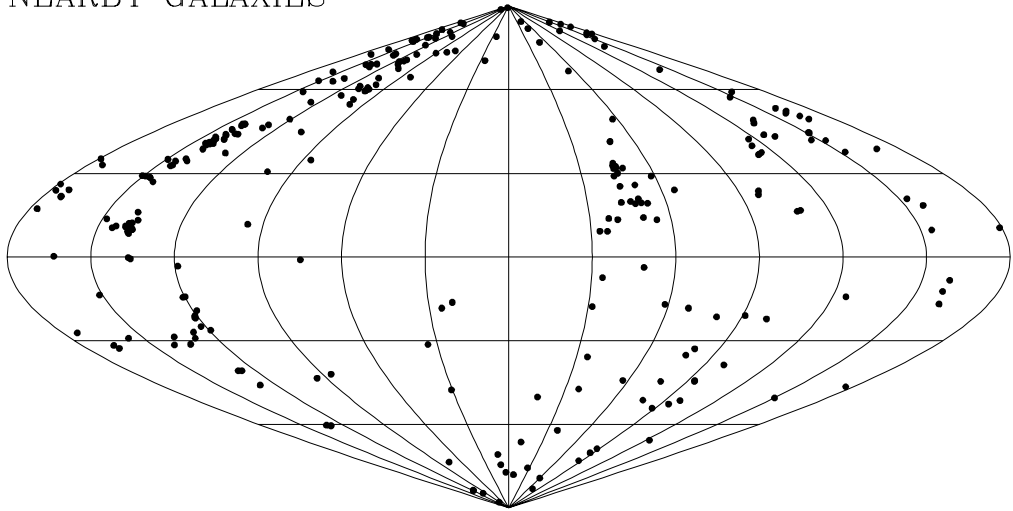

Fig. 4.— The distribution of 276 nearby galaxies in shown in galactic coordinates. The data were obtained from the CDS in Strasbourg (VII/161) and it is based on K.-H. Schmidt, T. Boller, & A. Priebe (1993, Astronomische Nachrichten, 313, no. 4, p. 189-231: Nearby galaxies. I - The catalogue). Notice the highly irregular distribution concentrated on the nearby Virgo cluster.

So much for the distributions which are observed. There are also distributions which can be theoretically inferred from what we know about the visible universe. One of those is very important for the debate: the distribution of high velocity neutron stars, This subject has been studied prior to the BATSE discovery by Paczyński (1990) and by Hartmann et al. (1990). If we assume that neutron stars are formed in supernova explosions in the galactic disk and ejected with some velocity in a random direction, then for velocities up to 630 km/sec a strong dipole or quadrupole anisotropy is always present if the galactic halo potential is assumed to be spherically symmetric. Podsiadlowski et al. (1995) found that if the potential of the extended dark halo is properly adjusted then the distribution of the very high velocity stars becomes isotropic at large distances. However, a strong concentration to the galactic center always remains, i.e. it is not possible to obtain a uniform density core of any radius. The very simple reason for the strong concentration is due to the combination of two facts: the starting points for the neutron star orbits are all within a few kiloparsecs of the galactic center as the galactic disk is there, and the gravitational potential of the inner galaxy must be centrally peaked to provide the observed flat rotation curve of the disk.

Let us summarize our findings. Among the many types of objects there are only two which satisfy the condition that they are random (not clustered) and isotropic: these are the distributions of the nearby stars at $\sim 100$ parsecs and the very distant extragalactic objects at $\sim 1$ Gigaparsec. Any intermediate distance scale reveals either the galactic structure or the local signature of the large scale structure of the universe. No exception is known! Of course this is nothing new. This was pointed out many times in the past (cf. van den Berg 1983, Paczyński 1991, and references therein). Now we are ready to consider the distribution of gamma-ray bursts.

In the pre-BATSE era it was well established that the bursters were distributed uniformly not only over the whole sky but also in distance (Atteia et al. 1987, Fenimore et al. 1992). Therefore, there was no way to choose which of the two distance scales was correct, galactic or cosmological. It was expected that BATSE, with its very high sensitivity, should resolve the distance puzzle. Its huge detectors were tested in balloon flights during which $\sim 60$ events should have been detected if their distribution followed the $N \sim F^{-1.5}$ law. Instead only three were recorded (Meegan et al. 1985). It was clear that the new detectors were so sensitive that they reached the "edge" of the burst distribution, but with only three events nothing could be said about the sky distribution. BATSE was designed to have not only very high sensitivity, but also to be able to locate the bursts with a precision of several degrees. Therefore, it should be very easy to check whether the weak bursts are concentrated towards the galactic plane or whether they are isotropic, thereby definitely solving the puzzle of their distance. Or so it seemed.

On September 23, 1991 the first results about the sky distribution of weak bursts from BATSE were presented at Annapolis, Maryland. Their sky distribution was isotropic, yet the distribution of intensities was approximately $N \sim F^{-0.8}$, a clear indication that the instrument could see beyond the region filled uniformly with bursters (Meegan et al. 1992). When the key viewgraphs were shown, a large fraction of the audience accepted them as clear evidence of the cosmological distance scale. Many participants of that historic conference changed their opinion from a galactic to cosmological origin because no concentration of the weak sources to the galactic plane was apparent in the data. Also, there was no concentration to the galactic center in direct conflict with any galactic halo distribution.

## 4. VARIOUS BURSTS AND THE GALACTIC CORONA

There is a number of different types of high energy bursts known. The best understood are X-ray bursts of type I, which are thermonuclear explosions on neutron stars accreting from a close binary companion star (Lewin & Joss 1981). A few dozen objects of this kind are known in our galaxy (cf. Fig. 6). X-ray bursts of type II are powered by some instability in the accretion flow onto a neutron star, which occasionally produces a type I burst as well (Hoffman et al. 1978). Only one object of this type is known: the Rapid Burster. The peak luminosity of all Type I X-ray bursts is within a factor of a few of $10^{38}$ $erg$ $s^{-1}$, their spectra are close to black body with an effective temperature of $\sim 2$ $keV$ at their peaks. They are believed to be old neutron stars, with an age of $10^8 - 10^9$ years.

There are three soft gamma repeaters known; all three are associated with supernova remnants: two in our galaxy, and one in the Large Magellanic Cloud, as shown in Fig. 6 (cf. Kulkarni et al. 1994, Murakami et al. 1994, Vasisht et al. 1994, and references therein). These three objects were suggested to be neutron stars with ultra strong magnetic fields, in the range $10^{14} - 10^{15}$ gauss (Duncan & Thompson 1992). Their peak luminosity is $\sim 10^{41} - 10^{42}$ $erg$ $s^{-1}$ (Kouveliotou et al. 1987, and references therein). The spectra look almost thermal, with most energy radiated at $\sim 30$ keV (Golenetskii et al. 1984, Paczyński 1992c, Fenimore et al. 1994), which is a factor $\sim 6$ higher than the peak of thermal spectra of X-ray bursts, and completely different from the broad non-thermal spectra of gamma-ray bursts, which extend to GeV energies (Hurley et al. 1994, and references therein). The soft gamma repeaters are super-Eddington events, $\sim 10^3 - 10^4$ times more luminous than X-ray bursts. As their name implies the soft gamma repeaters repeat.

There is also the unique March 5, 1979 event (GB790305b) which is related to the repeater SGR 0526-66 in the Large Magellanic Cloud, and which had a peak luminosity in excess of $10^{45}$ $erg$ $s^{-1}$, and a hard spectrum during its $\sim 0.2$ second duration (Mazets et al. 1982, Fenimore et al. 1995, and references therein). I shall refer to it by its popular name: 'March 5 event'.

The soft gamma repeaters are young: their age is $\sim 10^4$ years as judged from their supernova remnants. They are believed to be very high velocity neutron stars with magnetic fields of $\sim 10^{14} - 10^{15}$ gauss. This makes them distinctly different from the very high velocity neutron stars which are known to be radio pulsars: those have magnetic fields typical for all ordinary (non-millisecond) pulsars, i.e. $\sim 10^{12}$ gauss.

Finally, we have the gamma-ray bursts. The main idea behind their location in the galactic corona is the analogy with other bursters which have known distances. So, the crucial issue is: which of the many bursting types should we consider? Out of a large set of types there is only one which could be classified as a gamma-ray burst on the basis of its spectrum and time variability: the March 5 event. It is thought to be a young, very high velocity neutron star with an extraordinarily strong magnetic field. So, by direct inference we expect the gamma-ray bursters to have similar properties. However, their distribution is impossible to reconcile with an age of $\sim 10^4$ years (cf. Fig. 7). Hence to hold onto a galactic origin of gamma-ray bursts one has to make ad hoc assumption: (i) for some unknown reason the bursting activity which is present among the young very high velocity highly magnetized neutron stars like March 5 event, and stops after $\sim 10^4$ years, resumes after $10^7 - 10^8$ years. In addition (ii) it is necessary to have a broad distribution of the turn-on ages with a well adjusted shape so that the distribution of gamma-ray bursters would have an apparently uniform number density within $\sim 30$ kiloparsecs of the galactic center, which is absent in the distribution of the very high velocity neutron stars.

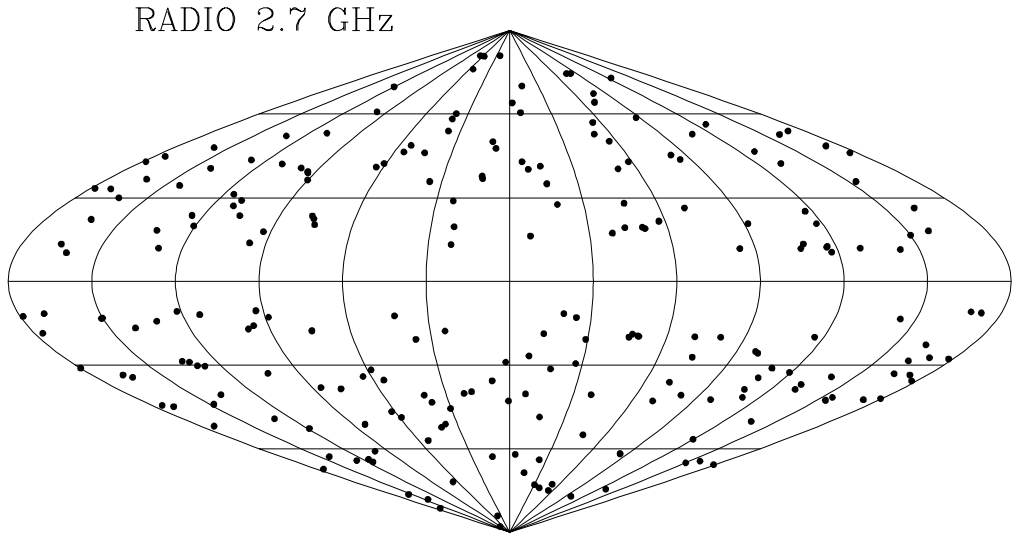

Fig. 5.— The distribution of the 233 strongest 2.7 GHz extragalactic radio sources is shown in galactic coordinates. The data were obtained from Wall & Peacock (1985). These sources are associated with very distant galaxies, and they are apparently distributed isotropically and randomly in the sky. Sources in the 'zone of avoidance' close to the galactic equator are not shown to avoid confusion with the large number of galactic sources.

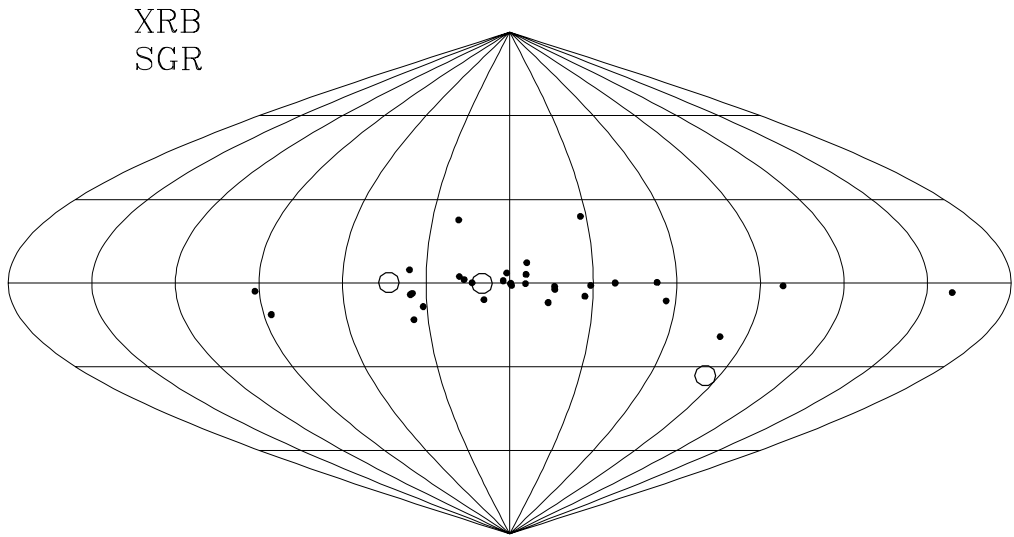

Fig. 6.— The distribution of the galactic X-ray bursters (XRB – filled circles) and the three known soft gamma repeaters (SGR – open circles) is shown in galactic coordinates. The data were obtained from J. van Paradijs (1995). Notice the strong concentration of sources toward the galactic plane and the galactic center.

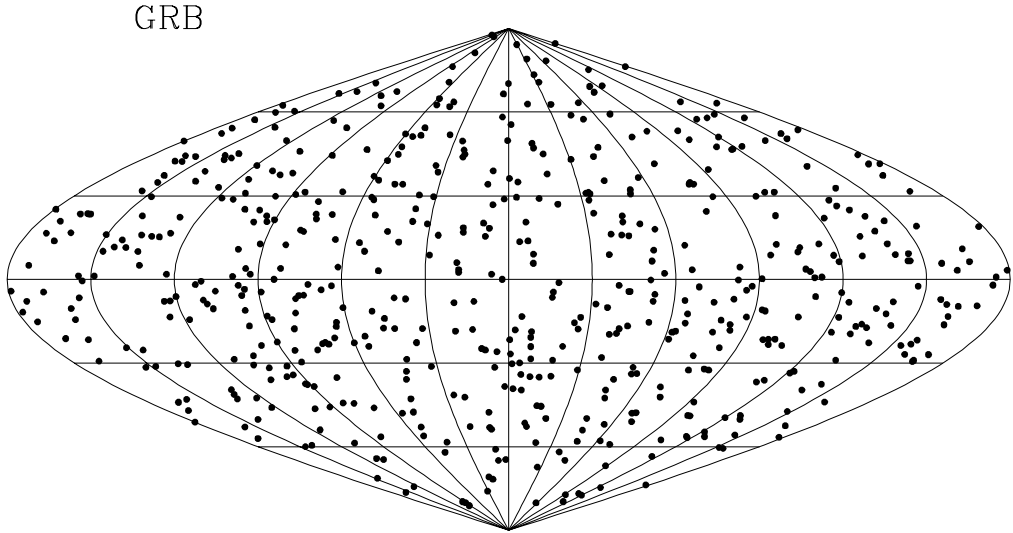

Fig. 7.— The distribution of the 585 gamma-ray bursts (GRB) from the second BATSE catalog is shown in galactic coordinates. The data were obtained from GRONEWS at Goddard Space Flight Center. Notice the random and isotropic distribution of sources over the sky.

The March 5 event was so intense that it could easily have been seen out to M31, and far beyond it. But there is no trace of M31 in the sky distribution of gamma-ray bursts. Therefore, we have to make another ad hoc assumption: (iii) gamma-ray bursts have to be a few orders of magnitude less intense than March 5 event. Also, (iv) we may have to make them stop bursting after some $3 \times 10^8$ years, otherwise those born in M31 would have reached our galaxy and a dipole towards M31 would be apparent in the data, but it is not. Or one could postulate that the burst emission is beamed only along the direction of their motion to hide the bursts from M31.

Let us estimate the energy required to sustain the activity of gamma-ray bursts. The rate observed by BATSE (allowing for its $\sim 40\%$ duty cycle) is $\sim 10^3$ bursts per year. If we assume that all soft gamma repeaters can also generate gamma-ray bursts then the birth rate of such objects is $3/10^4$ yr $\approx 3 \times 10^{-4}$ per year, unless there are many soft gamma repeaters which happen to be dormant at this time. Combining these two numbers we obtain $\sim 3 \times 10^6$ bursts per neutron stars, with $\sim 10^{42}$ ergs per burst (if they are at $\sim 100$ kiloparsec) for total energy of $\sim 3 \times 10^{48}$ ergs. This seems to be too staggering an energy to accept for those who prefer a galactic origin. So, another ad hoc assumption is made: (v) not only do the soft gamma repeaters produce gamma-ray bursts, but also the very high velocity radio pulsars. However, there is no measurable property of those pulsars (other than their high velocity) which makes them any different than their lower velocity counterparts. If the low velocity pulsars became gamma-ray bursters at the age of $\sim 10^8$ years then there would be a strong concentration of gamma-ray bursts to the galactic plane and to the galactic center. To prevent that from happening yet another ad hoc assumption has to be adopted: (vi) for some reason only the very high velocity radio pulsars, after proper aging, are capable of making gamma-ray bursts.

*So many ad hoc assumptions are necessary to make the coronal distribution of gamma-ray bursters consistent with the observations that my conclusion is: the coronal distribution is not a viable scientific proposition.*

## 5. THEORETICAL MODELS

When Ruderman (1975) presented his review of the newly discovered gamma-ray bursts at a Texas Symposium he noted there were more theories than the total number of bursts known at the time. In his recent reviews Nemiroff (1994a,b) compiled a list of over one hundred different theories. One does not have to know much about the subject to realize that if there is one correct theory of the bursts then all but one are wrong. One may continue this reasoning to note that if 99 out of 100 published theories are wrong then most likely all 100 are wrong. In other words, the multitude of proposals is the weakness and not the strength of the field. All theories of gamma-ray bursts ever presented were speculative, yet, they served a useful purpose by exploring so many diverse possibilities.

Over the years people got tired of the original unrestricted freedom and a consensus emerged that gamma-ray bursts were some energetic phenomenon on the surfaces of nearby neutron stars. There might have been a rational reason: the discovery of X-ray bursts. The story of X-ray bursts is one of the most spectacular success stories in modern astrophysics: a reasonably quantitative theory was developed within a year or two of their discovery, and it looks as good now (Lewin et al. 1995) as it did more than a decade ago (Lewin & Joss 1981). What made that success possible?

First, the sky distribution of X-ray bursts as shown in Fig. 6 instantly reveals their galactic origin and firmly sets their distance scale at about 8 kpc. Their spectra are thermal, very closely approximated by a Planck curve. Their peak luminosity is $\sim 10^{38}$ ergs per second, while the peak temperature is $\sim 2 \times 10^7$ K. Consequently the radius of the source is $\sim 10$ km. All this makes sense: 10 km is the radius of a neutron star, and $10^{38}$ erg/sec is the Eddington luminosity of a neutron star. A simple yet accurate model of a thermonuclear runaway followed.

The similarity between X-ray bursts and gamma-ray bursts is very limited. The X-ray bursts have thermal spectra that peak at about 5 keV. Gamma-ray bursts have very broad non-thermal spectra extending all the way from 1 keV to $10^7$ keV. X-ray bursts show a clear concentration to the galactic center, while gamma-ray bursts have an isotropic sky distribution, as shown in Fig. 7. X-ray bursts are known to repeat, while gamma-ray bursts either do not repeat or repeat very rarely (Petrosian & Efron 1995, Brainerd et al. 1995, and references therein).

The discovery of X-ray bursts contributed to the popularity of the galactic neutron star hypothesis for gamma-ray bursts, but very soon the latter acquired a life of its own. At any particular time certain types of gamma-ray burst models were fashionable. There were common ingredients to most of them: magnetized neutron stars and some source of free energy that was to be released very rapidly. There were starquakes, there were comets or nuggets of quark matter falling onto the surface of a neutron star, and there were phase transitions deep under the surface. And no consensus ever emerged as to the actual source of energy, or the physical process responsible for the observed emission of gamma-ray bursts.

Now we are witnessing a similar story with X-ray bursts replaced with the soft gamma repeaters as the alleged cousins of the gamma-ray bursts. The same type of reasoning which was claimed to provide evidence for a distance of $\sim 100$ parsecs to gamma-ray bursts is now supposed to provide evidence for a distance of $\sim 100$ kiloparsecs.

The possibility that gamma-ray bursters are in an extended galactic corona was proposed many times in the past (Fishman et al. 1978, Jennings & White 1980, Jennings 1982, Shklovski & Mitrofanov 1985, Atteia & Hurley 1986). Until recently no serious attempts were made to build physical models for the corona. Today the galactic corona is the site of choice for those who favor the galactic origin of gamma-ray bursters, relating them to the very high velocity neutron stars ejected from the galactic disk, and some models were suggested (Duncan & Thompson 1992, Li & Dermer 1992, Podsiadlowski et al. 1995, Lamb 1995).

There is also a number of models developed for gamma-ray bursters at cosmological distances, i.e. at $\sim 1$ Gpc (Paczyński 1991, 1992a, Mészáros et al. 1994, and references therein). Perhaps the colliding neutron stars became the most popular among them.

The relevance, or rather irrelevance of all models to the distance scale to gamma-ray bursters will be discussed in the next section.

## 6. DISCUSSION

Following BATSE's discovery the old galactic disk paradigm was practically eliminated as it was in direct conflict with the observed distribution. Two competing distance scales emerged: one was the galactic corona and the other was cosmological. The nature of the arguments leading to these two possibilities was (and still is) very different.

Ever since the extended corona was first proposed as the site for gamma-ray bursters the emphasis was on the radial extent and sphericity, hence the popularity of the very high velocity neutron stars which can go out as far as we please. But the really difficult problem with the coronal idea is not its outer extent, but the size of its inner core, the region over which the number density of the bursters has to be uniform, independent of location (Paczyński 1991, Hakkila et al. 1994, and references therein). In order not to see our 8 kiloparsec offset from the galactic center that core must be at least 30 kiloparsec in radius. In spite of the many papers written about the coronal models I am not aware of a satisfactory explanation of the extent of this uniform density region. The observed isotropy of the strong bursts (Atteia et al. 1987) combined with the observed radial uniformity of the PVO bursts (Fenimore et al. 1992, 1993, and references therein) demands that there is a region around us which has a constant number density of gamma-ray bursters. If that region is in our galaxy it must be at least 60 kiloparsecs in diameter. This is the most serious problem faced by any galactic corona model, as explained in section 4: far too many ad hoc assumptions have to be made to erase any trace of the galactic origin in the observed distribution of the bursts.

*The observed distribution is automatically satisfied if we adopt a cosmological distance scale: all objects detectable at cosmological distances are distributed isotropically in the sky, and the number density of all of them appears to decrease at sufficiently large distances as a consequence of the redshift. The uniform density region in almost any cosmological scenario is a few Gigaparsec in extent. The observed distribution is the most important argument for the cosmological distance scale to gamma-ray bursters.*

There are plenty of very schematic scenarios for any distance scale. As the sources are clearly non-thermal the diversity of ideas is limited only by the skill and imagination of theoreticians. This is not a unique situation. To the contrary: this is a typical situation. It is pretty much the same with all other non-thermal sources, like radio pulsars or quasars, even though we know the distance to those objects. Once we have to abandon the constraints of thermal equilibrium and are free to add magnetic fields, turbulence, relativistic particles, and other "high energy" ingredients, it is next to impossible to come up with a unique and highly quantitative description of what is going on. This is not to say that we should not try to invent new non-thermal models. In the process we find a large variety of interesting possibilities, and perhaps at some point we shall find a testable model which will turn out to be correct. I do not expect this to happen as long as we do not know for sure how far away the bursters are, and therefore we do not know what the energy requirements are.

Perhaps the most devastating argument against using models to determine the distance scale is the history of gamma-ray bursts. As far as I can see the galactic disk origin with its original distance scale of $\sim 100$ pc has been practically abandoned in spite of the fact that most people were convinced it was correct. The quality of models developed for the distance of $\sim 100$ kiloparsecs or $\sim 1$ Gigaparsec is not any higher than was the quality of the disk models: none is nearly good enough to be trusted. My conclusion is that the models of gamma-ray bursters as currently available are useless for distance determination. Each model is designed to work at some particular distance, be it 100 pc, 100 kpc, or 1 Gpc. In fact there are many scenarios for every possible distance. Even when the distance is finally established by model independent means we still shall not know which model, if any, is correct, as is still the case for quasars and for radio pulsars.

Consider the models developed to account for spectral features commonly known as 'cyclotron lines' – all distance scales are covered: $\sim 100$ parsecs (cf. Lamb 1992, and references therein), $\sim 100$ kiloparsecs (cf. Lamb 1995 and references therein), $\sim 1$ Gigaparsecs (Gould 1992, Stanek et al. 1993, Ulmer & Goodman 1995). A list of properties which are very important for the understanding of the bursts, but which cannot be used to establish the distance scale is long. In addition to spectral lines (if they exist) the list includes: the energetics, the rapid time variability, the repetition (if it exists), the absence of a spectral cutoff due to pair creation. All these can be accommodated at any distance with suitably speculative models.

Theoretical models are fun, and they may even be useful. But it is a major mistake to confuse 'evidence for' with 'consistent with'. All theoretical models of gamma-ray bursts ever developed were at best 'consistent with', or more likely 'not in conflict with', and fairly often 'in direct conflict with' the data. Not a single model was ever robust enough to provide 'evidence for'. As a community we should learn a lesson from the decade during which the bursters were firmly believed to be at $\sim 100$ parsecs, yet today we have no evidence that any of the processes proposed to justify that distance actually operates. Yet, the same mistake is being repeated today with the coronal as well as cosmological models - we have no evidence that any of those have anything to do with reality.

What can be done to establish the distance scale? Some steps have already been taken. The cosmological distance scale is suggested by the fact that the observed distribution is not only isotropic but also appears to be bounded. The latter effect comes about because the distant sources are redshifted. Therefore, we should look for observational consequences, of which at least two are promising. First, the redshift affects not only the wavelength of every photon but also the apparent duration of every burst (Paczyński 1992b, Piran 1992). Second, the spectra are not strict power laws, they are somewhat curved. Therefore, a redshifted spectrum should appear somewhat softer (Paczyński 1992b). The time dilation effect was apparent in some correlations, but the possibility of the redshift interpretation was not noticed for a while (Kouveliotou et al. 1992, Paciesas et al 1992). Recently, both effects were reported to be present in the data (Lestrade et al 1992, Norris et al. 1994, 1995, Nemiroff et al. 1994, Wijers & Paczyński 1994) at an approximately 4 $\sigma$ level of confidence. Some doubts remain (Mitrofanov et al. 1994, Band 1994, Fenimore & Bloom 1994), and it will take some time until a consensus is reached on the presence or absence of the effect, and on its interpretation. The case for cosmological redshift will be stronger when the two methods, time dilation and spectral softening, are shown to agree.

A common feature of all anti-cosmological arguments is that all of them are model dependent. The author(s) always have some specific model in mind when a claim is made that the model could not possibly give rise to a burst at a distance of a few Gigaparsecs. And usually the argument is correct, but only for that particular model, or for a particular set of models. There never was a proof that no model could possibly produce a burst at any specific distance scale. Also, there never was a model that could quantitatively demonstrate that it must produce a burst. *My conclusion is: we do not know what gamma-ray bursters are and we do not know what makes them burst. Therefore, if we want to determine their distance we must use a model independent method.*

There is a complication with the cosmological distance scale which should be mentioned here. All types of objects observed at cosmological distances, quasars, galaxies, and clusters of galaxies, are known to evolve: their typical luminosity and their number density observed at large redshifts are different than local. The population of gamma-ray bursters may also evolve. If there were no bursters in the early universe, then the observed distribution appears bounded because there are no bursters at large distances, and not because the redshift made the distant bursters very dim, and pushed them below the detection threshold. In this case the weakest BATSE events could be relatively nearby, and their redshifts may be too small to measure. What is interpreted as a redshift might instead be a consequence of evolution. This is a rather ugly possibility, but unfortunately it cannot be ruled out as long as we do not understand gamma-ray bursters.

As for the current galactic vs. cosmological controversy there is room for some optimism. Note that over the last few years the distance scale in the galactic scenario kept increasing from $\sim 100$ parsecs, to $\sim 1$ kiloparsec, and up to $\sim 100$ kiloparsecs today. In the same spirit the typical velocities went from $\sim 200$ $km\ s^{-1}$, to $\sim 400$ $km\ s^{-1}$, and up to $\sim 1,000$ $km\ s^{-1}$ today. The chances are that by the year 2,000 the distances will increase to $\sim 1$ Gigaparsec, the velocities will reach a fair fraction of the speed of light, and the only issue of the future debate with be the origin of those bursters. Presumably, some will claim that the bursters were ejected from our galaxy, pretty much like some claims that quasars were ejected from our galaxy.

## 7. CONCLUSIONS

There is strong evidence that gamma-ray bursters are at cosmological distances, but there is no rigorous proof. The most important question we can ask at this time is: how can one distance scale or another be proven? Note that all types of objects known to exist in our galaxy are always found in other galaxies as soon as the detectors are sensitive enough to uncover those objects at extragalactic distances. Therefore, we may have full confidence that if the known gamma-ray bursts are associated with our galaxy there should also be gamma-ray bursts associated with M31, which is the nearest giant spiral, pretty much like our own (Atteia & Hurley 1986). This issue cannot be resolved by BATSE, at least not yet (Hakkila et al. 1994), but a new experiment with $\sim 10$ times BATSE's sensitivity should resolve the issue (Harrison et al. 1994). If a concentration of very weak bursts towards M31 is detected we shall have to accept the fact that the gamma-ray bursters are in the corona of our Galaxy, establishing a new type of a distribution of astronomical objects. If no signature of M31 is found we shall have to accept the cosmological origin of the bursts.

I hope that all participants of this debate agree with my assessment of the decisive role of the experiment aimed at M31. Note, that extending the observed flux distribution of gamma-ray bursts to the level 10 times below the current BATSE limit is almost certain to be interesting in its own right: if the bursters are in the galactic corona we shall find where the corona is truncated; if they are cosmological we shall find them at larger redshifts and they may turn out to be useful for cosmology.

The current situation may be summarized as follows:

1. We do not know what gamma-ray bursters are or what makes them burst.

2. We have strong evidence that gamma-ray bursts are at cosmological distances.

3. We have no evidence for any other distance scale.

4. We have no evidence against a cosmological distance scale.

5. We do not have proof that the bursters are at cosmological distance.

6. An experiment $\sim 10$ times more sensitive than BATSE will determine the distance scale by comparing the number of the weakest bursts towards M31 with the number in any other direction.

7. An experiment providing $\sim 1$ arc second positions would greatly improve the likelihood that counterparts of gamma-ray bursters are finally detected at other wavelengths. This may be a critical step towards the determination of the nature of the bursters. A new interplanetary network would offer the best opportunity to achieve this goal.

It is a great pleasure for me to acknowledge many helpful comments and suggestions by W. H. G. Lewin, R. J. Nemiroff, J. van Paradijs, R. E. Rutledge, K. Z. Stanek, J. K. Wambsganss, and R. A. M. J. Wijers. It is also a great pleasure to acknowledge the warm hospitality by M. Doi, M. Fukugita, and S. Miyama during my visit to Japan, where this paper was written.

This work was supported with NASA grants NAG5-1901 and GRO/PFP-91-26.

---